\documentclass{easychair}

\usepackage{makeidx}
\makeindex

\usepackage{array}
\usepackage{multirow}

\newcommand{\xf}[1]{Figure~\ref{#1}}

\newcommand{\xs}[1]{Section~\ref{#1}}

\newcommand{\jini}{{Jini\index{Jini}}}

\newcommand{\gipsy}{{GIPSY\index{GIPSY}}}

\newcommand{\dms}{{DMS\index{DMS}}}
\newcommand{\jms}{{JMS\index{JMS}}}
\newcommand{\dmf}{{DMF\index{DMF}\index{Frameworks!DMF}}}

\newcommand{\lucid}{{Lucid\index{Lucid}}}

\newcommand{\jlucid}{{JLucid\index{JLucid}}}
\newcommand{\olucid}{{Objective Lucid\index{Tensor Lucid}}}

\newcommand{\flucid}{{Forensic Lucid\index{Forensic Lucid}}}

\newcommand{\jooip}{{JOOIP\index{JOOIP}}}

\newcommand{\cpp}{{C++\index{C++}}}

\newcommand{\file}[1]{\url{#1}\index{Files!#1}}

\newcommand{\api}[1]{\texttt{#1}\index{API!#1}}

\newcommand{\lucidL}[1]{{$\mathit{Lucid}$}($L$) }

		{}

\def\myvert{\raise 2.27pt \hbox{\vrule depth 0pt height 8pt width 0.2mm}}
\def\myarrow{\hspace*{0.43mm}%
             \raise 2.29pt\hbox{\vrule depth 0pt height 8pt width 0.16mm}%
             \hspace*{-0.32mm}%
             $\longrightarrow$
             \ %
             }

\begin{document}


\title{Towards Refactoring the DMF to Support {\jini} and JMS {\dms} in {\gipsy}}
\titlerunning{Uniform Invocation of Jini and JMS DMSs in GIPSY}

\author
{
	Yi Ji, Serguei A. Mokhov, and Joey Paquet\\
	Computer Science and Software Engineering\\
	Concordia University, Montreal, QC, Canada\\
	\url{{ji_yi,mokhov,paquet}@cse.concordia.ca}
}
\authorrunning{Ji, Mokhov, Paquet}


\maketitle

\begin{abstract}
In this paper we report on our re-engineering effort to refactor and unify
two somewhat disjoint Java distributed middleware technologies -- {\jini} and {\jms} --
used in the implementation of the Demand Migration System (DMS).
In doing so, we refactor their parent Demand Migration Framework (DMF), within
the General Intensional Programming System ({\gipsy}).
The complex Java-based {\gipsy} project is used to investigate on the
intensional and hybrid programming paradigms.
%
\end{abstract}

\hyphenation{In-te-r-o-pe-ra-bi-li-ty}




\section{Introduction}
\label{sect:introduction}

The {\gipsy} research prototype system \cite{gipsy} (see \xs{sect:background})
is a collection of replaceable Java components arranged primarily
into three main packages -- the collection of compilers for the programming
languages of interest (GIPC -- core -- Lucid-based intensional dialects \cite{lucid95}, and
hybrid dialects, primarily mixing Lucid code and Java code to various degrees),
runtime programming environment (RIPE) (currently a loose set of user interaction
components), and the run-time systems -- the general eduction engine (GEE).
We focus on some aspects of the latter in this paper, specifically its distributed
subcomponents that implement two instances of the demand-migration system (DMS)
using two distributed Java middleware technologies, namely Jini and JMS for
comparative studies of evaluation of hybrid programs to gain insight on the
technologies and various their parameters from programmability to scalability
metrics among others. Here we report some of our findings through development
and experiments.

\paragraph{Problem Statement}

One of the main reasons necessitating this study is the
hybrid intensional programming aspect the the GIPSY platform is there
to investigate among other things.
Lucid programs are naturally parallel and expressive \cite{lucid95,lucid85}
as well as context-oriented with contexts as first class
values \cite{wanphd06,tongxinmcthesis08}. Yet, in itself {\lucid} is rather
simple functional language for computation, and does not have rich I/O and
other support, so it should rely on the existing libraries and frameworks
when such needs arise leading the way to hybrid programming paradigms involving
a Lucid dialect and Java in our case.
Earlier (while still very valuable to the community, but relatively non-scalable and umaintainable)
solutions, were proposed and their enhancement with hybridification of Lucid and C \cite{glu1,glu2}
or later {\cpp} \cite{glu3}. Since GIPSY's architecture was first proposed and evolved in the
past 10 years to be extendable and component-based, hybrid prototype dialects emerged combinding
Java and Lucid -- {\jlucid} (Lucid program primarily calls only Java methods),
{\olucid} (Lucid to access to object properties of Java objects) \cite{mokhovmcthesis05} and later
{\jooip} (Java-based OO Intensional Programming) language \cite{aihuawu09} that enabled bidrectional
Lucid being able to access Java members and, at the same time, Java classes could contain fragments
written in Lucid in them. (Further work in programs involves extension of these in the form of
{\flucid} \cite{flucid-imf08} and MARFL \cite{marfl-context-secasa08}).
To support the evaluation of programs written in these hybrid dialects, the runtime (GEE)
of GIPSY has to scale to be able not only to locally compute light-weight Lucid fragments
locally, but also compute the hybrid ``heavy'' Java fragments -- given the latter can take
a lot of computation and I/O resources, and natural parallelization of Lucid, the hybrid
components are proposed to be evaluated distributively. Then the problem becomes which
distributed middleware technologies to pick.
As a proof of concept, the DMF was defined in Java and was implemented using two different
Java middleware technologies, {\jini} (by Vassev et al. \cite{dmf-plc05})
and {\jms} (Pourteymour \cite{dmf-pdpta07}). But those two were development in the
simulated prototype environment, relatively isolated from the core GIPSY project
and from each other.

\paragraph{Proposed Solution}

To enable consistent comparative studies of Jini and JMS in the GIPSY multi-tier
environment and hybrid language paradigms 
\cite{gipsy-multi-tier-sac09,gipsy-type-system-c3s2e09,bin-han-10} for points
of scalability \cite{ji-yi-mcthesis-2011}, usability, programmability, deployment,
and other aspects, we unify the two Java implementations of Jini and JMS DMS
under GEE and its multi-tier architecture and redefine the practical meaning
of certain of its components and do extensive testing of both. Hereafter, we
report on our experience in this regard.

\paragraph{Organization}

We proved the necessary background of the Java-based {\gipsy} project and its
distributed middleware technologies used in \xs{sect:background}.
We then describe the objectives of this work in \xs{sect:objectives}
and present the methodology and the approach in \xs{sect:methodology},
and finally we conclude in \xs{sect:conclusion}.


\section{Background}
\label{sect:background}

The General Intensional Programming System ({\gipsy}) provides a platform to
investigate the possibilities of the intensional programming~\cite{gipsy-multi-tier-secasa09}.
The intensional programming model, in the sense of Lucid, is a declarative and functional programming
language paradigm where the identifiers are evaluated in multidimensional context spaces.
The {\gipsy} compiler translates any flavor of intensional program into a source-language
independent GIPL program, and the {\gipsy} runtime system executes the GIPL program using
an evaluation model called eduction. In the demand-driven eduction model, an initial demand
requesting the value of a certain identifier is generated, and to consume this demand,
new demands are generated to request the values of the identifiers constituting the expression
defining the initial identifier, and similarly these demands further generate new demands
until eventually some of the demands are evaluated and propagated back in the chain of demands,
so that the identifiers whose value depend on them can be evaluated in turn,
and eventually the initial identifier is evaluated~\cite{gipsy-multi-tier-secasa09}.
This demand-driven eduction model naturally supports distributed
execution of intensional programs~\cite{glu2,glu3}.

The Demand Migration Framework ({\dmf}) for the {\gipsy} runtime system was proposed
for the distributed and demand-driven execution of intensional programs
by Vassev et al.~\cite{gipsy,%
vassev-mscthesis-05,%
dmf-plc05,%
dmf-cnsr05,%
dmf-pdpta07,%
mokhovmcthesis05}, and two principle Demand Migration Systems (DMS) as of this writing -- based
on {\jini} and {\jms} Java middleware technologies were provided to implement the DMF~\cite{dms-pdpta08,%
pourteymourmcthesis08%
}. The basic idea of the DMF is to provide a generic framework defining interface to
migrate/propagate demands among distributed demand generators and demand workers,
where the generators generate demands and the workers compute the demands.

A follow up work by Ji et al. \cite{ji-yi-mcthesis-2011} further streamlined
the code for unification while performing scalability studies for the the
Jini and JMS implementations of DMS in {\gipsy} (see \xs{sect:results}).

We briefly introduce the {\jini} DMS
and the {\jms} DMS architectures in the sections that follow.



\subsection{{\jini} DMS}
\label{sect:jini}

{\jini} is a Java-based and service-oriented middleware technology for building 
distributed systems consisting of {\jini} services and clients~\cite
{%
jini,%
jiniupnp,%
eermi%
}. 
Now officially known as {\em Apache River}~\cite{apache-river}, 
{\jini} defines a set of specifications and provides implementation
of several basic services such as lookup discovery, leasing and transaction services.
It also provides a Java implementation of the Tuple Space called the JavaSpace service
that enables distributed {\jini} clients to read, write and remove serialized 
Java objects stored in the shared object repository. JavaSpace supports
object persistence so that when restarted, the objects stored in the JavaSpace
can be recovered.

The {\jini} DMS of the {\gipsy} runtime system was implemented by Vassev
with his proposal of the first Demand Migration Framework (DMF)~\cite
{%
dmf-plc05,%
dmf-cnsr05%
}. As shown in \xf{fig:jini-architecture}, the {\jini} DMS consists of Demand Generators 
Demand Workers, {\jini} Transport Agent ({\jini} TA) and JavaSpace. 
The Demand Generators and Workers communicate among each other by sending and reading 
$demands$ into and from the JavaSpace with the aid of the {\jini} TA.
A $demand$ is a serialized Java object containing information 
for the evaluation of Lucid identifiers that require functional computation. 
A typical demand-migration process is as follows: 
\begin{enumerate}
	\item when parsing the abstract syntax tree of a hybrid Lucid program, 
				the Demand Generators encounter identifiers whose values depend on
				certain functional computations, so the Generators generate $pending$
				demands to request these computation and use the TA to send those
				$pending$ demands into the JavaSpace;
	\item the Demand Workers then use the TA to pick up the $pending$ demands from 
	      the JavaSpace and carry out the functional computations requested, 
	      and send the $computed$ demands back to the JavaSpace;
  \item the Demand Generators then use the TA to pick up the $computed$ demands 
        from the JavaSpace and retrieve the values of the identifiers.
\end{enumerate}

The {\jini} TA in \xf{fig:jini-architecture} consists of {\jini} TA proxies and
a {\jini} TA backend. Each Demand Generator or Worker can obtain a {\jini} 
TA proxy object using the {\jini} lookup discovery service, and uses this TA proxy to 
communicate with the remote {\jini} TA backend via remote method invocation.
To write a demand into the JavaSpace, once invoked remotely by the Demand Generator
or Worker, the {\jini} TA backend uses a local Demand Dispatcher object
to wrap the demands as JavaSpace entries and send the entries into the remote JavaSpace. 
The reference to the remote JavaSpace service held by the Demand Dispatcher 
is also looked up via the {\jini} lookup discovery service.
Similarly, once invoked remotely to read a demand from the JavaSpace, 
the {\jini} TA backend uses the Demand Dispatcher object to retrieve
the corresponding JavaSpace entry and unwrap it, and returns the
demand to the remote Demand Generator or Worker.

The separation of the {\jini} TA into proxies and the remote backend
it allows diverse {\jini} TA implementations to be 
integrated into the system without affecting existing components, and may increase
system availability by allowing the Demand Generators and Workers to connect
to any TA registered in the {\jini} lookup discovery service.
However, its disadvantages are that the additional remote method invocation 
is redundant that it increases communication cost and undermines performance.



\begin{figure}[htpb!]
	\centering
	\includegraphics[width=\columnwidth]{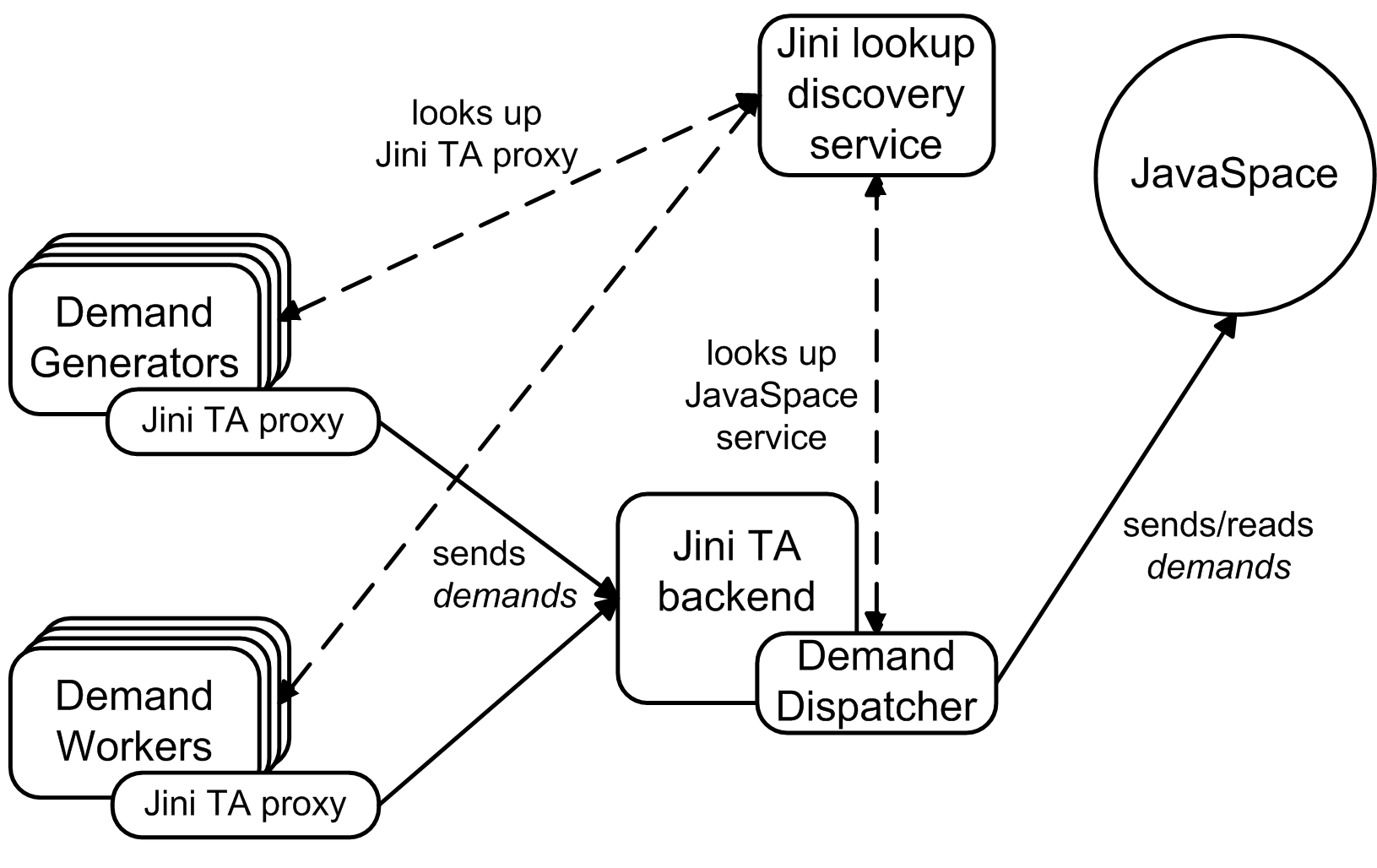}
	\caption{Jini DMS architecture}
	\label{fig:jini-architecture}
\end{figure}

%

\subsection{{\jms} DMS}
\label{sect:jms}

{\jms} is short for Java Message Service that is a Java-based and message-oriented 
middleware technology \cite{qosJMS,comparisonddsjms,jmstutorial}. 
It defines a set of API specifications and has several implementations
such Open Message Queue and JBoss Messaging. Basically a {\jms} implementation is a
{\jms} broker providing messaging services to its clients. The {\jms} clients fall into
two domains: message producer and consumer, and message publisher and subscriber.
The producer/consumer domain is for end-to-end messaging, meaning that each message
is sent by one producer, stored in the message queue in the broker, and received
by only one consumer; whereas the publisher/subscriber is for broadcast messaging,
in which each message is sent by one publisher, but can be received by multiple
subscribers. Compared to {\jini} JavaSpace, {\jms} broker provides additional
services to ensure availability and reliability such as delicate memory management, 
flow control, various acknowledgment models and better polished transactions. 
{\jms} also supports message persistence.

The {\jms} DMS was first implemented by Pourteymour~\cite
{%
dmf-pdpta07,%
dms-pdpta08,%
pourteymourmcthesis08%
} based on the message producer/consumer model. The alternative publisher/subscriber
model was not adopted because it does not allow newly registered subscribers to receive 
messages that were published before their registration, whereas the {\gipsy} must allow
workers to pick up $pending$ demands in the form of messages sent at anytime.
As shown in \xf{fig:jms-architecture}, the {\jms} DMS consists of Demand
Generators, Demand Workers, {\jms} TAs and the {\jms} broker service. Similar to the 
{\jini} DMS, the Demand Generators generate demands and the Workers compute 
the demands, and the demands are migrated among them via the TAs and the {\jms}
broker service. However, in this {\jms} DMS, to send a demand passed by the Demand
Generator or Worker, the {\jms} TAs wrap the demands into object messages and 
send them directly into the message queues managed by the broker; 
to read a demand, the {\jms} TAs directly read object messages
from the message queues, unwrap and return the demands
to the Demand Generator or Worker. Compared to the {\jini} DMS, the {\jms} DMS
has a simpler TA architecture without unnecessary communication cost, 
and it does not have a concrete Demand Dispatcher to wrap around the message queue
since the {\jms} API and broker was reckoned as the role of the Demand Dispatcher.
However, since the {\jms} TA in the {\jms} DMS is directly instantiated by the
Demand Generator or Worker, once the TA implementation is changed, the Java source
code of the existing components will be affected, which is inconvenient to add 
new TA implementations that are based on different middleware technologies.


%

\begin{figure}[htpb!]
	\centering
	\includegraphics[width=\columnwidth]{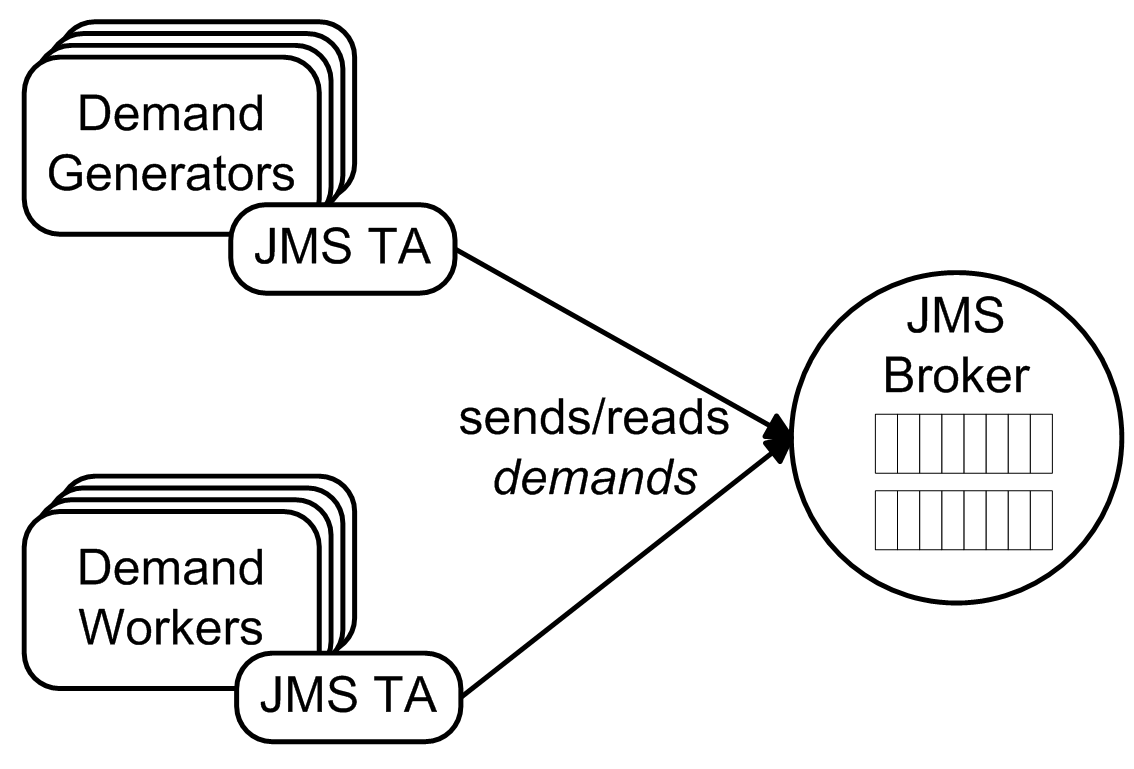}
	\caption{{\jms} DMS architecture}
	\label{fig:jms-architecture}
\end{figure}

%

\section{Methodology}
\label{sect:methodology}

We studies the original implementations of Jini and JMS DMS to be able
to run both consistently or even concurrently within one GIPSY computing network
instance. With this preliminary study we defined a number of required objectives
related to the actual refactoring, as well as identifying which technology is
more appropriate than the other and under which condition in the consistent
uniform setup; including common interfaces, glue, data structures and the like.

\subsection{Objectives}
\label{sect:objectives}


\begin{itemize}
	\item 
	Make {\jms}~\cite{jms} and {\jini}~\cite{jini} look similar, i.e. be a part of
	the same interchangeable framework's implementation.
	
	\item 
	Redefine the roles of Demand Dispatcher and the Transport Agent (see \xf{fig:generator-dispatcher})
	for the dispatcher to be more of decision maker and a scheduler for the generators, etc. rather than
	being attached to the demand store.
	
	\item
	Compare Jini and JMS vs JVM performance and scalability of computation.
	
	\item
	Compare programmability of the two APIs.
	
	\item
	Compare ease of deployment and startup of JMS and Jini from the same common point.
\end{itemize}

\begin{figure}[htpb!]
	\centering
	\includegraphics[width=\columnwidth]{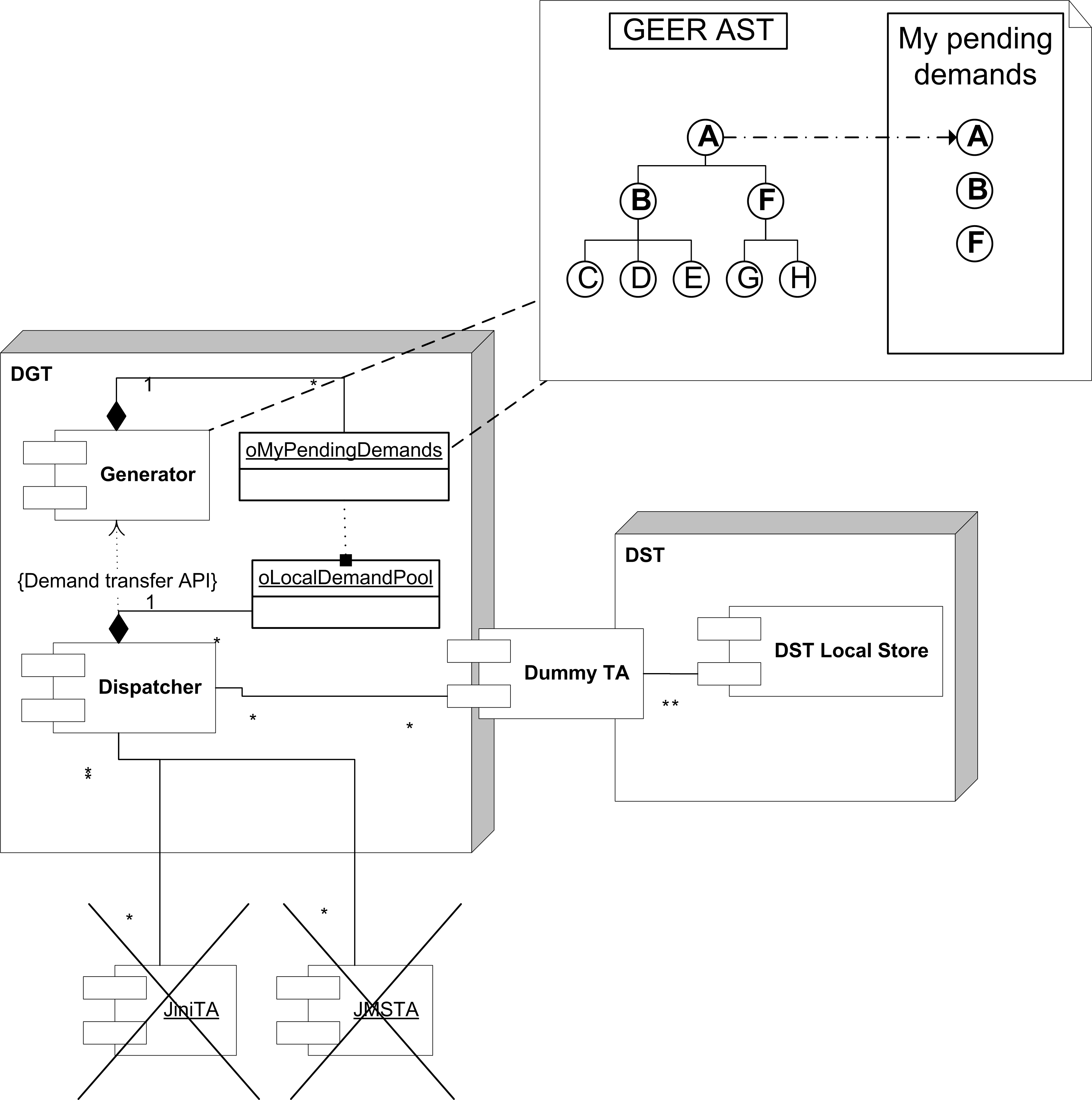}
	\caption{Demand Generator and Demand Dispatcher relationship}
	\label{fig:generator-dispatcher}
\end{figure}

\subsection{Making {\jini} and {\jms} similar}

	Before the refactoring process, the class diagram of the Jini and JMS DMSs
	is shown in \xf{fig:old-dms-class-diagram}. In this class diagram, the Jini 
	Demand Dispatcher communicates directly with the JavaSpace, and is used by
	the JTABackend that is remotely invoked by the JINITransportAgentProxy. 
	The JMSTA communicates with Message Queue directly and had no DemandDispatcher. 
	The JINITransportAgentProxy and the JMSTA inherit different interfaces as they
	expose too much middleware-dependent features, such as the UUid in {\jini} and the
	connection setup phase in {\jms}, and therefore each DMS has their own Demand 
	Generator and Worker since the Generator and Worker instantiate their TAs
	directly and use them in a middleware-dependent way.
	
	\begin{figure}[htpb!]
	\centering
	\includegraphics[width=\columnwidth]{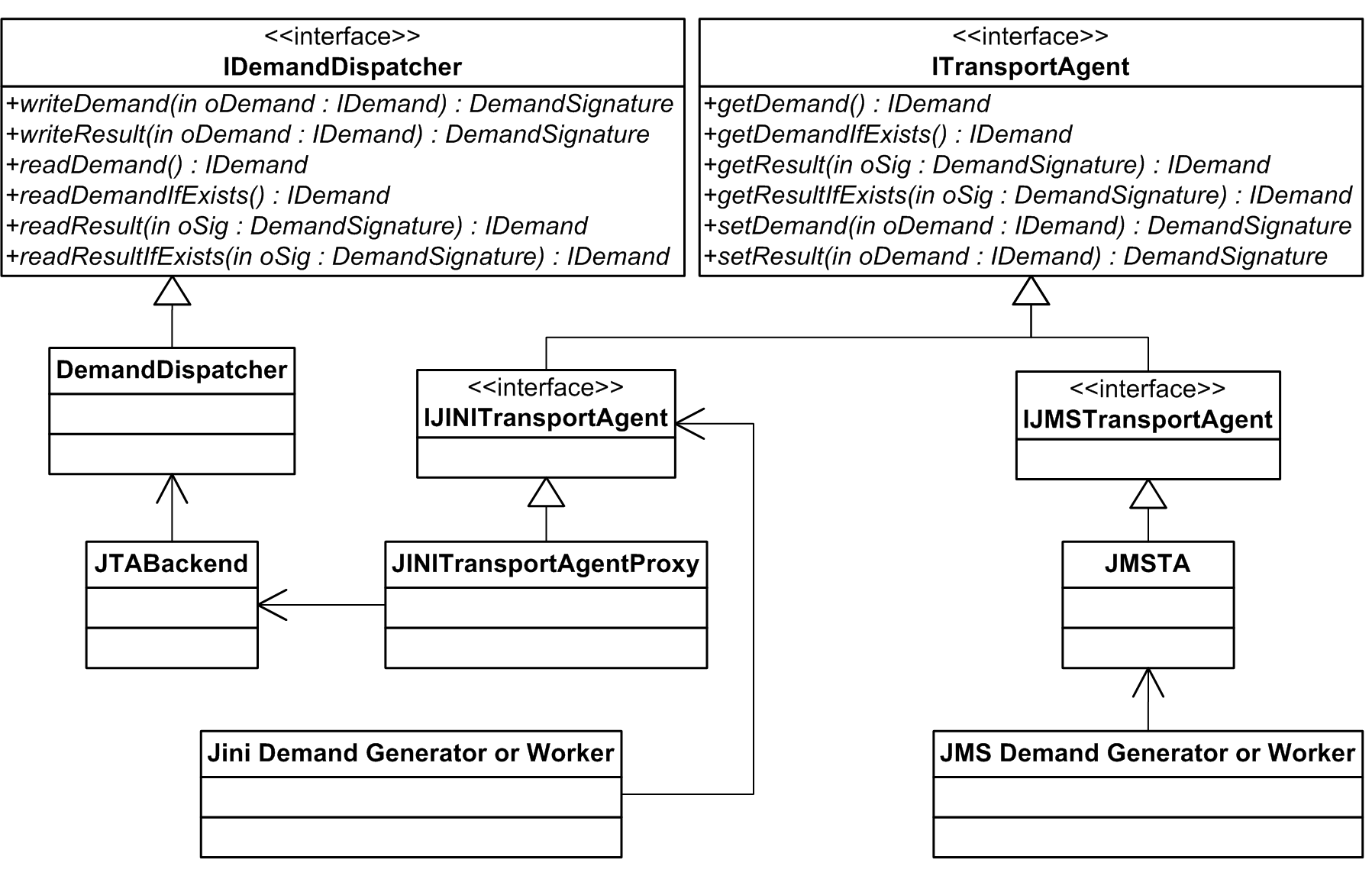}
	\caption{DMS class diagram before refactoring}
	\label{fig:old-dms-class-diagram}
	\end{figure}
	
	The refactoring began with creating a new class called JiniTA, moving the 
	original business logic of the DemandDispatcher into this JiniTA, and updating
	all the references to the original DemandDispatcher by pointing them to the
	new JiniTA. Then we encapsulated all the middleware-dependent logic, such
	as {\jms} connection setup and teardown, inside each TA implementation, and
	made them directly inherit the ITransportAgent interface. Having done so,
	we removed all the TA-implementation-dependent logic, such as TA instantiation,
	from the Demand Generators or Workers, so that they only reference ITransportAgent
	only. Then we used DemandDispatcher to delegate ITransportAgent, and replaced
	the usage of ITransportAgent with IDemandDispatcher inside the Demand Generator
	and Worker, so that the Demand Generators or Workers now talks to 
	IDemandDispatcher only and in the future we could easily add scheduling logic 
	into the DemandDispatcher to decide when, where and using what TA to send or receive
	demands. The class diagram of the DMS after our refactoring is shown in
	\xf{fig:new-dms-class-diagram}. We also unified the demand classes we use and
	separate them into subclasses due to their different purposes, such as
	procedural identifier evaluation, intensional identifier evaluation, runtime-resource
	acquisition and system management, as shown in \xf{fig:demand-tree}.
	
	\begin{figure}[htpb!]
	\centering
	\includegraphics[width=\columnwidth]{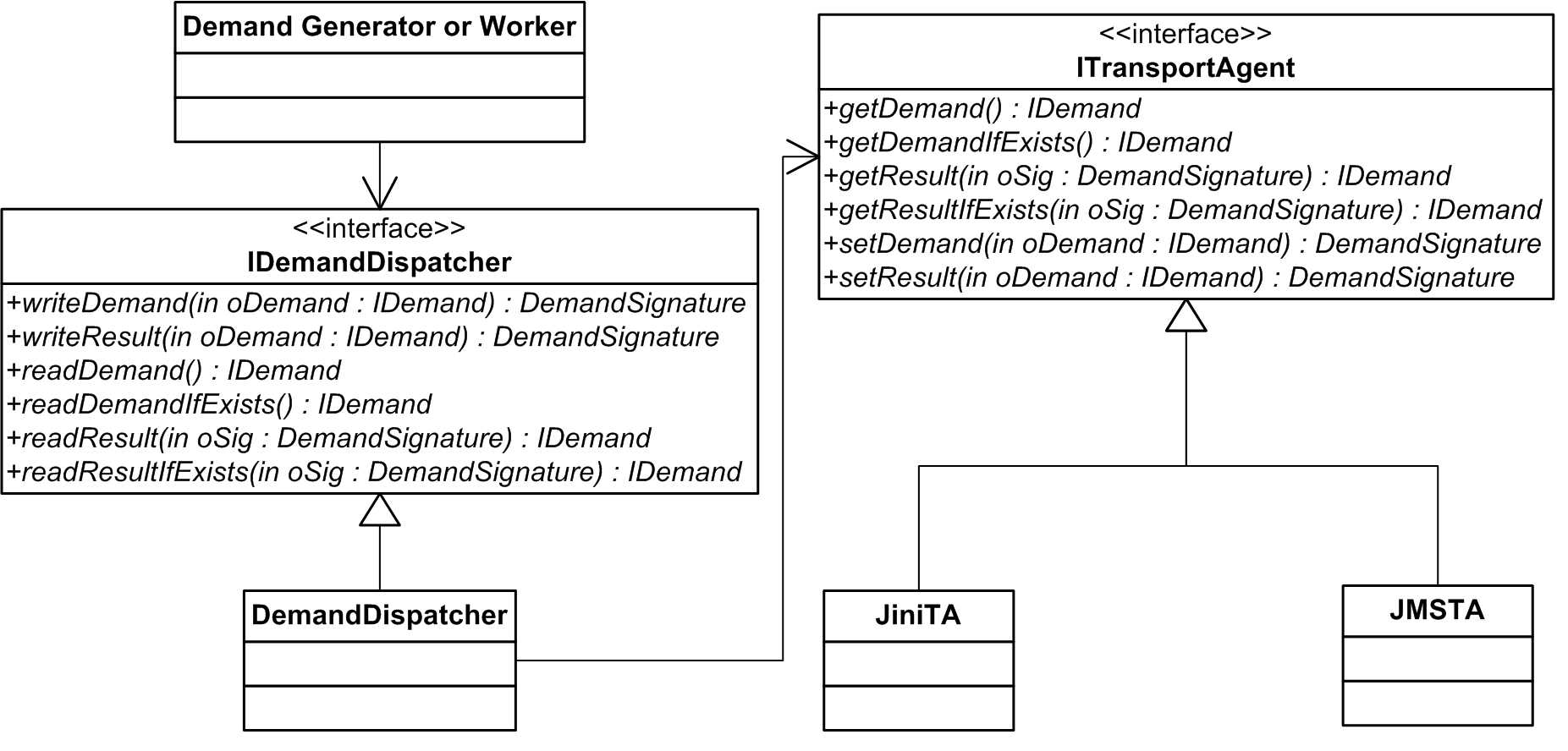}
	\caption{DMS class diagram after refactoring}
	\label{fig:new-dms-class-diagram}
	\end{figure}
	
	\begin{figure}[htpb!]
	\centering
	\includegraphics[width=\columnwidth]{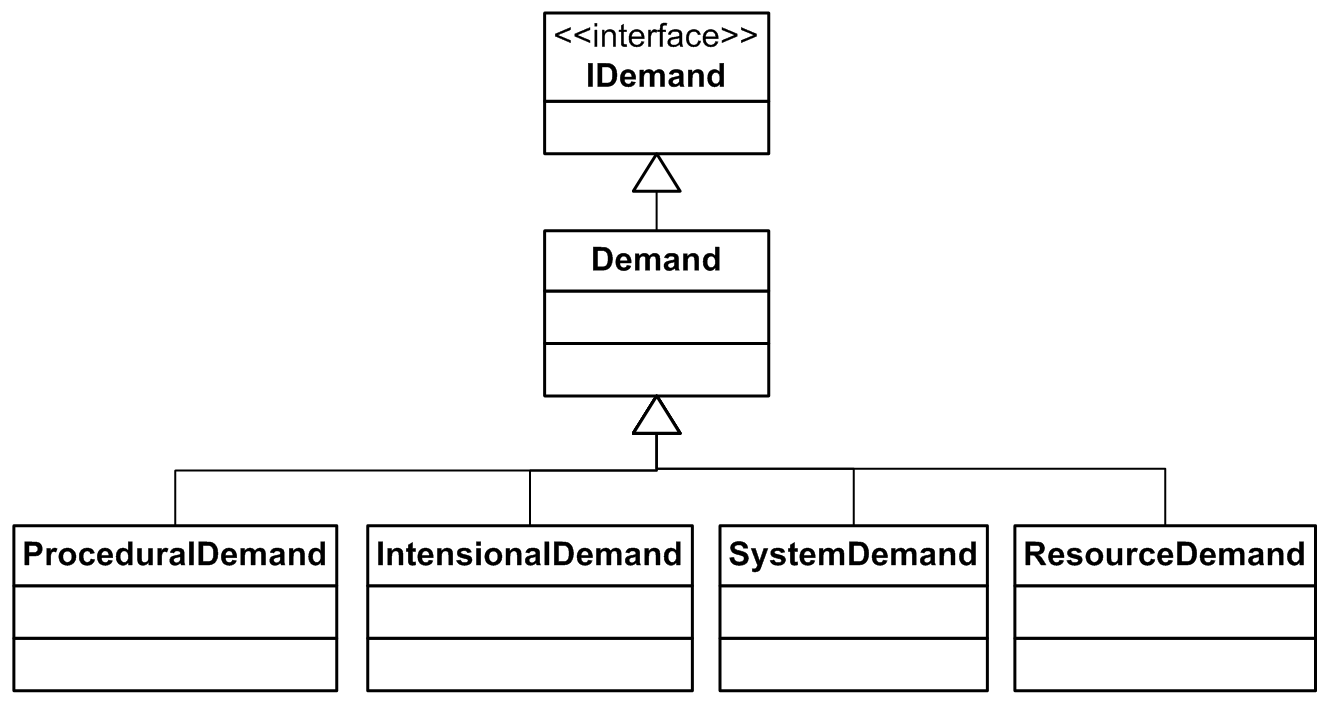}
	\caption{class diagram of demands}
	\label{fig:demand-tree}
	\end{figure}
	
	To ease the addition and switching of TA or Demand Dispatcher implementations,
	we use Java Reflection to instantiate each TA or Demand Dispatcher
	implementation by the name, where the name of the implementation is
	a character string passed via network or stored in files. In this way
	the two DMSs are interchangeable without changing the Java source code
	of the Demand Generator and the Demand Worker, so that we can use the 
	same set of Demand Generators and Demand Workers to a comparative
	study of the two DMSs. With proper testing and upon approval from all group members, 
	we committed the changes and cleaned up the affected code remained.
	

\section{Results}
\label{sect:results}

We compared the following three aspects of the Jini and JMS DMSs based on
our refactoring work.

	\subsection{Ease of programming}
	In our experience, the programming of {\jini} application requires the knowledge 
	of several separate but also correlated {\jini} concepts and services, 
	such as service lookup discovery, leasing, JavaSpace and transaction. 
	Such knowledge requires time and effort to collect, study and put into practice,
	especially if additional quality of service (QoS) is required. For example, {\jini} does
	not provide additional memory management besides the memory tuning available to the
	Java Virtual Machine (JVM), therefore if stronger memory management mechanism is required, 
	programmers will have to code their own {\jini} extension to enhance memory management.
	In contrast, {\jms} is a mainstream middleware technology and has relatively integrated
	services, more QoS choices, well managed documentation and easily understandable tutorials, 
	which is easier for programmers to learn and put into practice.
		
	\subsection{Ease of deployment}
	{\jini} requires only a set of .jar and configuration files to start its services, therefore
	has a light weight (the size of all .jar files is less than 4 MB) and can be easily
	deployed across different platforms, as long as Java is installed in the machines. In contrast,
	{\jms} requires platform-dependent executable binary files, such as .exe files in Windows,
	to start and manage the broker service, therefore it has a larger size (in the case of 32-bit Windows
	version, the size of all executable files and .jar files is around 20 MB), and requires those
	plat-form dependent executable files to be deployed across different platform.
	
	\subsection{Runtime issues}
	We tested that when the {\jini} DMSs was storing increasing amount of demands, 
	the {\jini} DMS would run out of memory and crash in the end, even if the data persistence feature
	was turned on. This shows the storage capacity of the {\jini} JavaSpace is constrained by its
	memory, and it provides no mechanism to prevent JVM crash. In contrast, the {\jms} DMS
	with message persistence turned on can swap persistent messages into its persistent storage, such
	as files, to reduce its memory usage once the memory usage exceeds certain threshold.
	Therefore when the {\gipsy} runtime system is facing increasing amount of demands, the {\jms} DMS
	with message persistence is a better choice when the system availability is the dominant concern.
	
	However, when comparing system performance in the sense of throughput of concurrent Pi
	calculation demands when the demands were generated by a multi-threaded Generator and computed
	by multiple Demand Workers distributed in different computers with maximumly two Workers
	per computer, we found that as the number Workers increases, the {\jini} DMS provided a higher throughput than
	the {\jms} DMS, and the {\jms} DMS reached its throughput saturation when there were 10 Workers deployed.
	The test result is shown in \xf{fig:scalability-load-test-2}, and the test was perform in a lab
	with computers with the same hardware and operating system environment shown in \xf{tab:test-environment}.
	These computers and their corresponding switch ports have 100 Mbps maximum speeds, 
	with each machine connected to one switch port and all of them on the same subnet and VLAN.
	This test shows that in the sense of throughput, the {\jini} DMS is a better choice than the {\jms} DMS.
	We also found that for each Demand Generator or Worker connection, the {\jini} DMS uses one thread
	to handle each connection, whereas the {\jms} DMS uses two threads. 
	
	All the above differences show that when deployed in managed network, {\jini} is more suitable for DMS 
	requiring high throughput but low memory storage and low reliability, whereas {\jms} is suitable for DMS 
	requiring high reliability and availability.
	
	\begin{figure}[htpb!]
	\centering
	\includegraphics[width=\columnwidth]{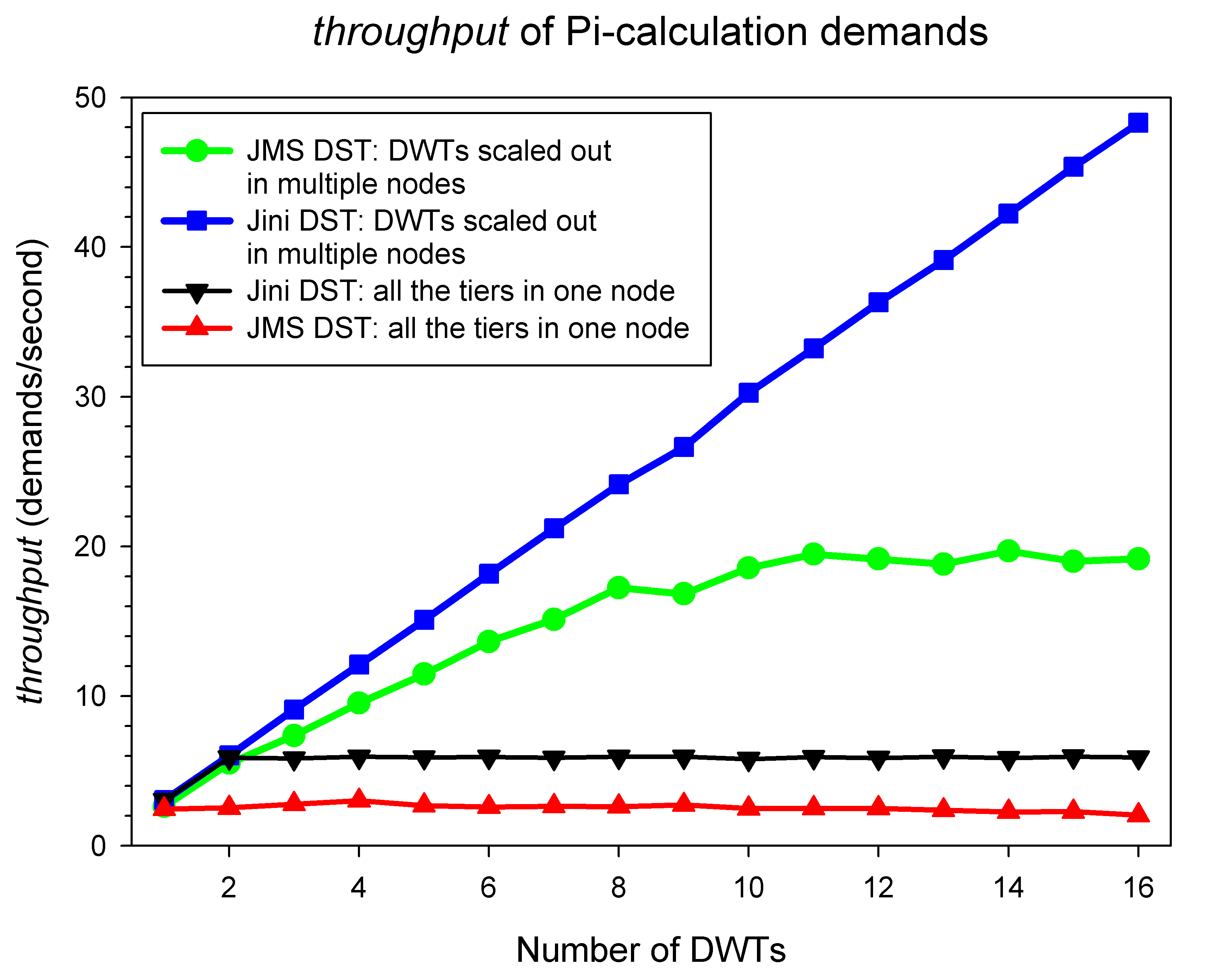}
	\caption{throughput of Pi calculations of the two DMSs}
	\label{fig:scalability-load-test-2}
	\end{figure}
	
	\begin{table*}[htbp]
		\centering
		\caption{Hardware and operating system environment}
		\label{tab:test-environment}
		\begin{tabular}{ | l | p{8cm} | }
		\hline
		OS Name                    & Microsoft Windows 7 Enterprise \\ \hline
		Version                    & Version	6.1.7600 Build 7600 \\ \hline
		System Type                & X86-based PC \\ \hline
		\multirow{2}{*}{Processor} & Intel(R) Core(TM)2 CPU 6300 @ 1.86GHz, 1862 MHz, 2 Core(s), 2 Logical Processor(s) \\ \hline
		Installed RAM	             & 2.00 GB \\ \hline
		Total RAM	                 & 2.00 GB \\ \hline
		Available RAM	             & 1.06 GB \\ \hline
		Total Virtual Memory       & 4.00 GB \\ \hline
		Available Virtual Memory   & 2.58 GB \\ \hline
		Page File Space	           & 2.00 GB \\ \hline
  	\end{tabular}
	\end{table*}
	

\section{Conclusion}
\label{sect:conclusion}

We have successfully did the POC integration of the
two middleware technologies implementations based
on {\jini} and {\jms} available to the GIPSY run-time
system. In the future work we plan to continue
refactoring and cleaning up the other technologies
within GIPSY to work together in unison.
It is evident, that Jini appears to be easier to work with for
development and deployment, but its memory-bound scalability is
more problematic than that of JMS, so JMS-based implementation is
general more reliable, but Jini DMS offers higher throughput
over JMS.
For in-depth results of the initial study on various
scalability metrics pleas refer to \cite{ji-yi-mcthesis-2011}.
Some significant redesign was also necessary to make the two
implementations work together consistently, with a potential
payoff any new, better or worse, implementations for comparative
studies like we did, will be much more manageable.

\subsection{Future work}

There is a lot of work to be done; our immediate future attention will be
the item (1) below to expand our distributed testing environment followed
by other proposed items:

\begin{enumerate}
	\item Various JVMs in Linux and MacOS X distributed testing environments and clusters
	\item Comparative study for Web Services-based implementation
	\item Long-running distributed computation processes (e.g. MARF pattern recognition pipeline with very large data set over GIPSY)
	\item Expand the architecture to mobile Java platforms
\end{enumerate}


\subsection{Acknowledgment}

This work in part is supported by NSERC and the Faculty of Engineering and
Computer Science, Concordia University, Montreal, QC, Canada.
We thank reviewers for their constructive reviews and feedback.

\bibliographystyle{plain}
\bibliography{unifying-refactoring-jini-jms-dms}

\begin{thebibliography}{10}

\bibitem{jiniupnp}
J.~Allard, V.~Chinta, S.~Gundala, and G.~G.~Richard III.
\newblock {JINI} meets {UPnP}: An architecture for {JINI/UPnP}
  interoperability.
\newblock In {\em Proceedings of the 2003 International Symposium on
  Applications and the Internet 2003}. SAINT, 2003.

\bibitem{apache-river}
{Apache River Community}.
\newblock Apache {River}.
\newblock [online], 2010.
\newblock \url{http://incubator.apache.org/river/index.html}.

\bibitem{lucid95}
Edward~A. Ashcroft, Anthony~A. Faustini, Rangaswamy Jagannathan, and William~W.
  Wadge.
\newblock {\em Multidimensional Programming}.
\newblock Oxford University Press, London, February 1995.
\newblock {ISBN}: 978-0195075977.

\bibitem{qosJMS}
S.~Chen and P.~Greenfield.
\newblock {QoS} evaluation of {JMS}: An empirical approach.
\newblock In {\em Proceedings of the 37th Hawaii International Conference on
  System Sciences}, 2004.

\bibitem{eclipse}
{Eclipse contributors} et~al.
\newblock {Eclipse Platform}.
\newblock eclipse.org, 2000--2011.
\newblock \url{http://www.eclipse.org}, last viewed February 2010.

\bibitem{eermi}
R.~Eggen and M.~Eggen.
\newblock Efficiency of distributed parallel processing using {Java RMI},
  sockets, and {CORBA}.
\newblock In {\em Proceedings of the 2001 International Conference on Parallel
  and Distributed Processing Techniques and Applications (PDPTA'01)}. PDPTA,
  June 2001.

\bibitem{bin-han-10}
Bin Han.
\newblock Towards a multi-tier runtime system for {GIPSY}.
\newblock Master's thesis, Department of Computer Science and Software
  Engineering, Concordia University, Montreal, Canada, 2010.

\bibitem{jmstutorial}
M.~Hapner, R.~Burridge, R.~Sharma, J.~Fialli, and K.~Stout.
\newblock {\em Java(TM) Message Service API Tutorial and Reference}.
\newblock Prentice Hall PTR, 2002.
\newblock {ISBN} 0201784726.

\bibitem{glu1}
Raganswamy Jagannathan and Chris Dodd.
\newblock {GLU} programmer's guide.
\newblock Technical report, SRI International, Menlo Park, California, 1996.

\bibitem{glu2}
Raganswamy Jagannathan, Chris Dodd, and Iskender Agi.
\newblock {GLU}: A high-level system for granular data-parallel programming.
\newblock In {\em Concurrency: Practice and Experience}, volume~1, pages
  63--83, 1997.

\bibitem{ji-yi-mcthesis-2011}
Yi~Ji.
\newblock Scalability evaluation of the {GIPSY} runtime system.
\newblock Master's thesis, Department of Computer Science and Software
  Engineering, Concordia University, Montreal, Canada, March 2011.

\bibitem{jini}
{Jini Community}.
\newblock Jini network technology.
\newblock [online], September 2007.
\newblock \url{http://java.sun.com/developer/products/jini/index.jsp}.

\bibitem{comparisonddsjms}
R.~Joshi.
\newblock A comparison and mapping of {Data Distribution Service (DDS)} and
  {Java Message Service (JMS)}.
\newblock Real-Time Innovations, Inc., 2006.

\bibitem{mokhovcvs}
Serguei~A. Mokhov.
\newblock {GIPSY}: {CVS} service on {Newton}, a crush guide, June 2003.
\newblock
  \url{http://newton.cs.concordia.ca/~gipsy/cgi-bin/viewcvs.cgi/*checkout*/resources/doc/presentations/cvs.pdf?rev=HEAD}.

\bibitem{mokhovmcthesis05}
Serguei~A. Mokhov.
\newblock Towards hybrid intensional programming with {JLucid}, {Objective
  Lucid}, and {General Imperative Compiler Framework} in the {GIPSY}.
\newblock Master's thesis, Department of Computer Science and Software
  Engineering, Concordia University, Montreal, Canada, October 2005.
\newblock {ISBN} 0494102934; online at \url{http://arxiv.org/abs/0907.2640}.

\bibitem{marfl-context-secasa08}
Serguei~A. Mokhov.
\newblock Towards syntax and semantics of hierarchical contexts in multimedia
  processing applications using {MARFL}.
\newblock In {\em Proceedings of the 32nd Annual {IEEE} International Computer
  Software and Applications Conference ({COMPSAC})}, pages 1288--1294, Turku,
  Finland, July 2008. IEEE Computer Society.

\bibitem{flucid-imf08}
Serguei~A. Mokhov, Joey Paquet, and Mourad Debbabi.
\newblock Formally specifying operational semantics and language constructs of
  {Forensic Lucid}.
\newblock In Oliver G{\"o}bel, Sandra Frings, Detlef G{\"u}nther, Jens Nedon,
  and Dirk Schadt, editors, {\em Proceedings of the IT Incident Management and
  IT Forensics (IMF'08)}, LNI140, pages 197--216. GI, September 2008.

\bibitem{gipsy-type-system-c3s2e09}
Serguei~A. Mokhov, Joey Paquet, and Xin Tong.
\newblock A type system for hybrid intensional-imperative programming support
  in {GIPSY}.
\newblock In {\em Proceedings of C3S2E'09}, pages 101--107, New York, NY, USA,
  May 2009. ACM.

\bibitem{glu3}
Nikolaos~S. Papaspyrou and Ioannis~T. Kassios.
\newblock {GLU\#} embedded in {C++}: a marriage between multidimensional and
  object-oriented programming.
\newblock {\em Softw., Pract. Exper.}, 34(7):609--630, 2004.

\bibitem{gipsy-multi-tier-secasa09}
Joey Paquet.
\newblock Distributed eductive execution of hybrid intensional programs.
\newblock In {\em Proceedings of the 33rd Annual IEEE International Computer
  Software and Applications Conference ({COMPSAC}'09)}, pages 218--224,
  Seattle, Washington, USA, July 2009. IEEE Computer Society.

\bibitem{gipsy-multi-tier-sac09}
Joey Paquet.
\newblock A multi-tier architecture for the distributed eductive execution of
  hybrid intensional programs.
\newblock Unpublished, 2009.

\bibitem{dmf-pdpta07}
Amir~Hossein Pourteymour, Emil Vassev, and Joey Paquet.
\newblock Towards a new demand-driven message-oriented middleware in {GIPSY}.
\newblock In {\em Proceedings of {PDPTA 2007}}, pages 91--97, Las Vegas, USA,
  June 2007. {PDPTA}, CSREA Press.

\bibitem{dms-pdpta08}
Amir~Hossein Pourteymour, Emil Vassev, and Joey Paquet.
\newblock Design and implementation of demand migration systems in {GIPSY}.
\newblock In {\em Proceedings of {PDPTA 2009}}. CSREA Press, June 2008.

\bibitem{pourteymourmcthesis08}
Amir~Hossein Pouteymour.
\newblock Comparative study of {Demand Migration Framework} implementation
  using {JMS} and {Jini}.
\newblock Master's thesis, Department of Computer Science and Software
  Engineering, Concordia University, Montreal, Canada, September 2008.

\bibitem{jms}
{Sun Microsystems, Inc.}
\newblock {Java Message Service (JMS)}.
\newblock [online], September 2007.
\newblock \url{http://java.sun.com/products/jms/}.

\bibitem{gipsy}
{The GIPSY Research and Development Group}.
\newblock The {General Intensional Programming System (GIPSY)} project.
\newblock Department of Computer Science and Software Engineering, Concordia
  University, Montreal, Canada, 2002--2011.
\newblock \url{http://newton.cs.concordia.ca/~gipsy/}, last viewed February
  2010.

\bibitem{tongxinmcthesis08}
Xin Tong.
\newblock Design and implementation of context calculus in the {GIPSY}.
\newblock Master's thesis, Department of Computer Science and Software
  Engineering, Concordia University, Montreal, Canada, April 2008.

\bibitem{dmf-cnsr05}
Emil Vassev and Joey Paquet.
\newblock A general architecture for demand migration in a demand-driven
  execution engine in a heterogeneous and distributed environment.
\newblock In {\em Proceedings of the 3rd Annual Communication Networks and
  Services Research Conference ({CNSR} 2005)}, pages 176--182. IEEE Computer
  Society, May 2005.

\bibitem{dmf-plc05}
Emil Vassev and Joey Paquet.
\newblock A generic framework for migrating demands in the {GIPSY}'s
  demand-driven execution engine.
\newblock In {\em Proceedings of the 2005 International Conference on
  Programming Languages and Compilers ({PLC} 2005)}, pages 29--35. CSREA Press,
  June 2005.

\bibitem{vassev-mscthesis-05}
Emil~Iordanov Vassev.
\newblock General architecture for demand migration in the {GIPSY}
  demand-driven execution engine.
\newblock Master's thesis, Department of Computer Science and Software
  Engineering, Concordia University, Montreal, Canada, June 2005.
\newblock {ISBN} 0494102969.

\bibitem{lucid85}
William~W. Wadge and Edward~A. Ashcroft.
\newblock {\em Lucid, the Dataflow Programming Language}.
\newblock Academic Press, London, 1985.

\bibitem{wanphd06}
Kaiyu Wan.
\newblock {\em Lucx: {Lucid} Enriched with Context}.
\newblock PhD thesis, Department of Computer Science and Software Engineering,
  Concordia University, Montreal, Canada, 2006.

\bibitem{aihuawu09}
Ai~Hua Wu.
\newblock {\em {OO-IP} Hybrid Language Design and a Framework Approach to the
  {GIPC}}.
\newblock PhD thesis, Department of Computer Science and Software Engineering,
  Concordia University, Montreal, Canada, 2009.

\end{thebibliography}

\appendix

\section{How to run the {\jini} DMS and the {\jms} DMS in the {\gipsy} project}
This documents introduces the steps to deploy and start the {\gipsy} runtime system
in 32-bit Windows.

\begin{enumerate}
	\item check out the GIPSY project \cite{mokhovcvs}. The project is an Eclipse
	project so you can simply imported into Eclipse IDE for Java Developers \cite{eclipse}.
	However, before you import it into Eclipse, remember to stop Eclipse from
	cleaning the bin folder as all the DMS executables are there. For example, in the case of 
	Eclipse Helios, once the IDE is open, go to Windows->Preferences->Java->Compiler->Building, 
	and uncheck the option ``Scrub output folders when cleaning projects'', and go back
	to the menu bar, click Project, and uncheck the option ``Build Automatically''
	so that you will need to build the project manually. Then you can import the {\gipsy}
	project and build it manually.
	
	\item Once the {\gipsy} project is built, you can go to the project's home folder,
	then go to the directory \\bin\\multitier\\.
	
	\item If you want to start {\jini} DMS, open StartGMTNode.config, and set the value of
	the property gipsy.GEE.multitier.Node.DSTConfigs as ../jini/DST.config
	
	\item If you want to start {\jms} DMS, open StartGMTNode.config, and set the value of
	the property gipsy.GEE.multitier.Node.DSTConfigs as ../jms/DST.config
	
	\item Double click startGMTNode.bat.
	
\end{enumerate}

\section{How to run the {\jini} DMS in the {\gipsy} project}
\label{sect:howto}

This document introduces the steps to compile and run the {\jini} code in the GIPSY project in
Windows XP. To use this guide, readers are required to have basic Java and Eclipse experience,
basic Jini knowledge (for example, service, lookup, JavaSpace, etc), and the basic understanding
of the GIPSY project structure.

\begin{enumerate}
	\item 
Get and install the appropriate software, and import the GIPSY project \cite{mokhovcvs}.
NOTE: If the software below cannot be found online, please check the GIPSY tools repository
	\begin{enumerate}
		\item 
		JDK 6 update 14 or later (\url{http://java.sun.com/javase/downloads/index.jsp}). It is better to set
		the \api{JAVA\_HOME} environment variable.
		\item 
		Eclipse IDE for Java Developers~\cite{eclipse}. Unpack the IDE, open
		it and import the GIPSY project from the GIPSY CVS into the Eclipse.
		\item 
		Jini Technology Starter Kit v2.1~\cite{jini}. The
		installation should require no administrator accounts. The term \api{JINI\_HOME} would be used in
		this manual to refer the installation directory.
	\end{enumerate}

	\item 
Compile the Jini code of the GIPSY project
NOTE: Point 1 and 2 should be done already in the GIPSY project when you get it.
	\begin{enumerate}
		\item 
		Copy the \file{jini-core.jar} and \file{jini-ext.jar} from \file{JINI\_HOME/lib} into the \file{gipsy/lib}
		\item 
		Configure the project Build Path by adding the above jars inside lib through ``Add JARs''
		\item 
		Build the project automatically or manually.
	\end{enumerate}

	\item 
Start the Jini service
	\begin{enumerate}
		\item 
Go to \file{JINI\_HOME/installverify}, and double-click the Launch All shortcut. The shortcut should
launch a service window and a Service Browser window.
		\item 
In the Service Browser window, click the ``Registrar'', and select the registrar representing
your computer. Then there would be 6 services appear in the ``Matching Services'' area,
including JavaSpace05, LookupDiscoveryService, ServiceRegistrar and TransactionManager.
		\item 
Leave the two windows open and do not touch them unless you want to shut down all the
services.
	\end{enumerate}

	\item 
Run the code requiring only JavaSpace.
NOTE: Currently there are two groups of Jini scenarios. The first one consists of
\api{DemandDispatcher}, \api{DemandDispatcherClient} and the \api{DemandDispatcherAgent} under the
\api{gipsy.GEE.IDP} package. The second scenario consists of the \api{gipsy.GEE.IDP.DemandDispatcher}, and
the DGT class in the \api{gipsy.GEE.IDP.DemandGenerator.simulator} package, and the \api{Worker} class in
the \api{gipsy.GEE.IDP.DemandGenerator.simulator.jini} package.

	\begin{enumerate}
		\item 
		Open Eclipse and run the classes within the same groups mentioned above with the \api{main()} method.
	\end{enumerate}

\item
Run the code requiring both JavaSpace and RMI
	\begin{enumerate}
		\item 
		Use command window to go to the \file{gipsy/bin} directory.
		\item 
		Open the \file{startJiniHTTPServer.bat} in edit mode, and check if all the paths are correct.
		\item 
		Double-click the \file{startJiniHTTPServer.bat}.
		\item 
		Double-click the \file{startJiniRMID.bat}.
		\item 
		Open Eclipse and run the\\\api{gipsy.GEE.IDP.DemandGenerator.jini.rmi.JINITransportAgent}.
		and the \api{gipsy.GEE.IDP.DemandGenerator.simulator.jini.WorkerJTA}, and the \api{DGT} in
		the \api{gipsy.GEE.IDP.DemandGenerator.simulator} package.
		
		Note:
		
		Please make sure that the \file{startJiniHTTPServer.bat} and the \file{startJiniRMID.bat} are in the \file{gipsy/bin}
		folder. If they are missing, please refer to the following content. Please make sure the settings in
		the content are consistent with your own JRE directories as shown in \xf{fig:startHTTPServer}
		and \xf{fig:startJiniRMID}.

\begin{figure*}[htpb]%
\hrule
\scriptsize
\begin{verbatim}

set RUNTIME_JAR="D:\Program Files\Java\jre6\lib\rt.jar"
set JINIHOME_BACKSLASH="D:\Program Files\jini2_1"
set JINI_CLASSPATH=.;%RUNTIME_JAR%;%JINIHOME_BACKSLASH%\lib\jini-core.jar;%JINIHOME_BAC
KSLASH%\lib\jini-ext.jar;%JINIHOME_BACKSLASH%\lib\reggie.jar;%JINIHOME_BACKSLASH%\lib-d
l\reggie-dl.jar;%JINIHOME_BACKSLASH%\lib\mahalo.jar;%JINIHOME_BACKSLASH%\lib-dl\mahalo
-dl.jar;%JINIHOME_BACKSLASH%\lib\outrigger.jar;%JINIHOME_BACKSLASH%\lib-dl\outrigger-dl.j
ar;%JINIHOME_BACKSLASH%\lib\tools.jar;%JINIHOME_BACKSLASH%\lib\sun-util.jar;

java -jar -classpath %JINI_CLASSPATH% %JINIHOME_BACKSLASH%\lib\tools.jar -port 8085 -dir .
-verbose
\end{verbatim}
\normalsize
\hrule
\caption{\texttt{startHTTPServer.bat}}%
\label{fig:startHTTPServer}%
\end{figure*}

\begin{figure*}[htpb!]%
\hrule
\begin{verbatim}

rmid -J-Djava.security.policy=gipsy/GEE/IDP/config/jini.policy
\end{verbatim}
\hrule
\caption{\texttt{startJiniRMID.bat}}%
\label{fig:startJiniRMID}%
\end{figure*}

	\end{enumerate}
\end{enumerate}
%
%

%
%

\printindex

\end{document}